\documentclass[pdftex]{pasj02} 
\usepackage[switch,mathlines]{lineno} 
\usepackage{natbib}  
\usepackage{url}
\usepackage{subfigure}
\usepackage{siunitx}
\usepackage{booktabs}

\jyear{2025}
\Received{2025 April 9}
\Accepted{2025 July 10}

\begin{document} 
\title{SILVIA: Ultra-precision formation flying demonstration for space-based interferometry}

\author{
Takahiro~\textsc{Ito},\altaffilmark{1}\orcid{0000-0003-1491-1940}
Kiwamu~\textsc{Izumi},\altaffilmark{1}\orcid{0000-0003-3405-8334}
Isao~\textsc{Kawano},\altaffilmark{1}
Ikkoh~\textsc{Funaki},\altaffilmark{1}\orcid{0000-0001-9193-1967}
Shuichi~\textsc{Sato},\altaffilmark{2}\orcid{0000-0001-5560-5224}
Tomotada~\textsc{Akutsu},\altaffilmark{3}\orcid{0000-0003-0733-7530}
Kentaro~\textsc{Komori},\altaffilmark{4,5}\orcid{0000-0002-4092-9602}
Mitsuru~\textsc{Musha},\altaffilmark{6}\orcid{0000-0001-7741-4584}
Yuta~\textsc{Michimura},\altaffilmark{4,7}\orcid{0000-0002-2218-4002}
Satoshi~\textsc{Satoh},\altaffilmark{8}\orcid{0000-0003-4017-5123}
Takuya~\textsc{Iwaki},\altaffilmark{9}\orcid{0000-0001-6809-6035}
Kentaro~\textsc{Yokota},\altaffilmark{9}\orcid{0000-0001-8160-1446}
Kenta~\textsc{Goto},\altaffilmark{9}\orcid{0009-0002-2704-7901}
Katsumi~\textsc{Furukawa},\altaffilmark{1}\orcid{0009-0008-1598-8281}
Taro~\textsc{Matsuo},\altaffilmark{10,11}\orcid{0009-0003-7304-7512}
Toshihiro~\textsc{Tsuzuki},\altaffilmark{3}\orcid{0000-0002-8342-8314}
Katsuhiko~\textsc{Yamada},\altaffilmark{12}\orcid{0000-0001-8717-1133}
Takahiro~\textsc{Sasaki},\altaffilmark{13}\orcid{0000-0002-2646-7173}
Taisei~\textsc{Nishishita},\altaffilmark{13}\orcid{0009-0002-7586-5589}
Yuki~\textsc{Matsumoto},\altaffilmark{13}
Chikako~\textsc{Hirose},\altaffilmark{9}\orcid{0000-0002-7675-4285}
Wataru~\textsc{Torii},\altaffilmark{1}\orcid{0009-0004-1004-1348}
Satoshi~\textsc{Ikari},\altaffilmark{14}\orcid{0000-0002-0467-768X}
Koji~\textsc{Nagano},\altaffilmark{15,16}\orcid{0000-0001-6686-1637}
Masaki~\textsc{Ando},\altaffilmark{5,4}\orcid{0000-0002-8865-9998}
Seiji~\textsc{Kawamura},\altaffilmark{11}\orcid{0000-0002-0841-4529}
Hidehiro~\textsc{Kaneda},\altaffilmark{11}\orcid{0000-0001-6879-1556}
Shinsuke~\textsc{Takeuchi},\altaffilmark{1}\orcid{0000-0002-4771-6175}
and
Shinichiro~\textsc{Sakai}\altaffilmark{1}\orcid{0009-0007-4685-3622}
}

\altaffiltext{1}{Institute of Space and Astronautical Science, Japan Aerospace Exploration Agency, Sagamihara, Kanagawa 252-5210, Japan}
\altaffiltext{2}{Department of Science and Engineering, Hosei University, Koganei, Tokyo 184-8584, Japan}
\altaffiltext{3}{National Astronomical Observatory of Japan, Mitaka, Tokyo 181-8588, Japan}
\altaffiltext{4}{Research Center for the Early Universe (RESCEU), Graduate School of Science, The University of Tokyo, Bunkyo, Tokyo 113-0033, Japan}
\altaffiltext{5}{Department of Physics, The University of Tokyo, Bunkyo, Tokyo 113-0033, Japan}
\altaffiltext{6}{Institute for Laser Science, University of Electro-Communications, Chofu, Tokyo 182-8585, Japan}
\altaffiltext{7}{Kavli Institute for the Physics and Mathematics of the Universe (Kavli IPMU), WPI, UTIAS, The University of Tokyo, Kashiwa, Chiba 277-8568, Japan}
\altaffiltext{8}{Department of Mechanical Engineering, The University of Osaka, Suita, Osaka 565-0871, Japan }
\altaffiltext{9}{Research and Development Directorate, Japan Aerospace Exploration Agency, Sagamihara, Kanagawa 252-5210, Japan}
\altaffiltext{10}{Graduate School of Science, The University of Osaka, Toyonaka, Osaka 560-0043, Japan}
\altaffiltext{11}{Graduate School of Science, Nagoya University, Nagoya, Aichi 464-8602, Japan}
\altaffiltext{12}{Osaka Metropolitan University, Sakai, Osaka 599-8531, Japan}
\altaffiltext{13}{Research and Development Directorate, Japan Aerospace Exploration Agency, Tsukuba, Ibaraki 305-8505, Japan}
\altaffiltext{14}{Department of Aeronautics and Astronautics, The University of Tokyo, Bunkyo, Tokyo 113-8656, Japan}
\altaffiltext{15}{Institute for Multidisciplinary Sciences, Yokohama National University, Yokohama, Kanagawa 240-8501, Japan}
\altaffiltext{16}{LQUOM, Inc., Yokohama, Kanagawa 240-8501, Japan}
\email{ito.takahiro@jaxa.jp; izumi.kiwamu@jaxa.jp}



\maketitle

\begin{abstract}
We propose SILVIA (Space Interferometer Laboratory Voyaging towards Innovative Applications), a mission concept designed to demonstrate ultra-precision formation flying between three spacecraft separated by 100 m. SILVIA aims to achieve sub-micrometer precision in relative distance control by integrating spacecraft sensors, laser interferometry, low-thrust and low-noise micro-propulsion for real-time measurement and control of distances and relative orientations between spacecraft. A 100-meter-scale mission in a near-circular low Earth orbit has been identified as an ideal, cost-effective setting for demonstrating SILVIA, as this configuration maintains a good balance between small relative perturbations and low risk for collision. This mission will fill the current technology gap towards future missions, including gravitational wave observatories such as DECIGO (DECihertz Interferometer Gravitational wave Observatory), designed to detect the primordial gravitational wave background, and high-contrast nulling infrared interferometers like LIFE (Large Interferometer for Exoplanets), designed for direct imaging of thermal emissions from nearby terrestrial planet candidates. The mission concept and its key technologies are outlined, paving the way for the next generation of high-precision space-based observatories.
\end{abstract}


\section{Introduction}
\label{sec:Intro}
Since the first direct detection of gravitational waves in 2015 \citep{GW150914}, over 250 events have been reported by the LIGO-Virgo-KAGRA collaboration, shedding light on the nature of black holes and neutron stars. Similarly, since the discovery of a planet around a Sun-like star in 1995 \citep{MayorQueloz1995}, more than 5,000 exoplanets have been identified, offering new insights into the potential for life beyond Earth. These advancements highlight the exciting possibilities opened up by innovative observational technologies, uncovering the mysteries of the universe from black holes to distant exoplanets.

Multiple gravitational wave detections have enabled statistical analyses of binary properties, including merger rates, mass and spin distributions, and the redshift evolution of mergers. However, current detections remain limited to stellar-mass binaries with component masses below roughly 100 ${\rm M_\odot}$, and the sky localization of binary mergers is not precise enough to reliably identify their host galaxies with the current generation of gravitational wave detectors \citep{ObsScenario}. To address these challenges, extending the observational band to lower frequencies below approximately \SI{10}{Hz} offers a promising path forward.

Space-based detectors, such as LISA (Laser Interferometer Space Antenna) \citep{LISA2017} and DECIGO (DECihertz Interferometer Gravitational wave Observatory) \citep{DECIGO2001,DECIGO2021}, are particularly well suited for observing low-frequency gravitational waves. These detectors avoid the seismic noise and gravity gradient noise that dominate at low frequencies on Earth and allow test masses to behave as nearly ideal free masses. DECIGO is a concept focusing on the decihertz range, a critical band that enables the detection and characterization of intermediate-mass binary black hole mergers \citep{Matsubayashi2004}, provides early alerts and precise localizations for binary mergers \citep{Liu2020}, and probes the early universe prior to the cosmic microwave background, an era inaccessible to electromagnetic observations \citep{Calcagni2021}. Achieving the strain sensitivity necessary for these scientific objectives requires constructing interferometers with optical cavities between multiple spacecraft to suppress the shot noise of laser light. However, maintaining the relative displacement between spacecraft within the linear operational range of these interferometers, typically less than the laser wavelength of approximately \SI{1}{\micro m}, presents a significant technical challenge \citep{Nagano2021,Izumi2021,Sugimoto2024}. Additionally, the relative alignment between spacecraft must be controlled to the microradian level to ensure that the beam spot displacement remains smaller than the beam diameter to maintain the interferometer visibility~\citep{Michimura2025}. These challenges distinguish DECIGO from LISA, which adopts an optical transponder scheme for interferometry between spacecraft and implements constellation flying. 

This type of formation flying is not only essential for space-based gravitational wave observatories but also holds significant importance for nulling infrared interferometry aimed at exoplanet exploration. One major objective in exoplanet research is to investigate the surface environments of potentially habitable planets and to search for biosignatures in their atmospheres through spectroscopy. Direct imaging of exoplanets can generally follow two paths: observing visible-wavelength reflected light or detecting mid-infrared thermal emission. Reflected light provides information about the surface composition of a planet, while infrared thermal emission reveals its atmospheric composition and vertical temperature structure \citep[e.g.][]{Fujii2018}. Spectroscopic observations in both the visible and infrared ranges thus provide complementary insights into the characterization of habitable planet candidates.

Atmospheric spectroscopy of these candidates requires the faint light of the planet to be separated from the bright glare of its host star and to suppress the stellar signal. Detecting reflected light in the visible range requires high-contrast coronagraphic imaging on a large space telescope. On the other hand, detecting thermal emission in the mid-infrared demands a stellar interferometer with a baseline of approximately \SI{100}{m} to spatially resolve the planet from its host star. In addition, achieving the required suppression of stellar light at a contrast level of $10^{-6}$ in the mid-infrared on nulling interferometers necessitates stabilizing the optical path difference (OPD) between the collected beams to within \SI{1}{nm} (rms) and controlling the relative tilt to a sub-milli arcsecond level over durations of several tens of hours \citep[e.g.][]{Lay2004,Matsuo2023}. These stringent requirements are met through a combination of high-precision formation flying and internal optical compensation using a delay line system and a tip-tilt mirror. The spectral characterization of thermal emissions from potentially habitable exoplanets was selected as one of the key science themes in ESA's Voyage 2050, driving momentum behind the development of the LIFE (Large Interferometer For Exoplanets) mission concept \citep{Quanz2022} within the astronomy, planetary science, and astrobiology communities.

Precise formation flying is essential for both DECIGO and LIFE, as it provides the baseline stability needed to enable micro- to nanometer-level optical path control, using test-mass and laser frequency actuators in the case of DECIGO, and internal delay lines in the case of LIFE. However, the precision of onboard navigation and control in past missions was limited to the centimeter scale. The first successful demonstration of autonomous formation flying, rendezvous, and docking was attained by ETS-VII (Engineering Test Satellite VII) \citep{Kawano2001result, OK03} in 1998, whose control accuracy was on the order of centimeters at a docking phase. Later, in 2010, PRISMA (Prototype Research Instruments and Space Mission technology Advancement) \citep{PVB09} demonstrated autonomous formation flying and rendezvous between two satellites at distances ranging from 100 to 2000~m, achieving control accuracy from centimeter to meter levels. While the centimeter- to submeter-order accuracy of formation control is sufficient for some astronomical applications (e.g., the Formation Flight All Sky Telescope (FFAST) mission \citep{tsunemi2008high} and Exo-Starshade (Exo-S) mission \citep{starshade2015}), others can require the higher precision and accuracy. The ongoing PROBA-3 (Project for On Board Autonomy-3) mission \citep{LlorenteEtc13, Penin2020proba3} (launched in 2024) aims to form a solar coronagraph using two satellites approximately \SI{150}{m} apart with a relative displacement accuracy under a sub-millimeter and pointing accuracy at the arc-second level. Until June 2025, the two PROBA-3 spacecraft demonstrated a precise formation flying at a distance of 150 meters with an accuracy of millimeters (range direction) and sub-millimeters (lateral direction) \footnote{Proba-3 achieves precise formation flying, \url{https://www.esa.int/Enabling_Support/Space_Engineering_Technology/Proba-3_achieves_precise_formation_flying} [Accessed June 23, 2025]}, and took the first images of the solar corona \footnote{Proba-3’s first artificial solar eclipse, \url{https://www.esa.int/Enabling_Support/Space_Engineering_Technology/Proba-3/Proba-3_s_first_artificial_solar_eclipse} [Accessed June 23, 2025]}.

In addition, high-precision formation flying technologies have been developed by utilizing small satellites and CubeSats. The VISORS (VIrtual Super Optics Reconfigurable Swarm) mission \citep{Lightsey2018cgj,Koenig2023visors} aims to image the solar corona in high resolution by forming a distributed telescope with two 6U CubeSats \SI{40}{m} apart. The successful science observation requires a centimeter-to-millimeter-level accuracy of formation geometry for 10 seconds. The STARI (STarlight Acquisition and Reflection toward Interferometry) mission concept \citep{monnier2024stari} aims to advance critical system-level technologies for space interferometry, controlling a three-dimensional CubeSat formation to the few mm-level and reflecting starlight over tens to hundreds of meters from one spacecraft to another. Additionally, several recent mission concepts \citep{MDDL24, IkariEtc21, Kruger2024Starling} have utilized or plan to utilize more than two spacecraft for autonomous formation flying. Among them, the SEIRIOS (Space Experiment of InfraRed Interferometric Observation Satellite) \citep{IkariEtc21, Matsuo2022} mission is scheduled to launch around 2030. SEIRIOS consists of an approximately \SI{40}{kg} nanosatellite and two 6U CubeSats and aims to obtain interferometric fringes from stellar objects as the first stellar interferometer in space by controlling the distance between the two CubeSats centered on the nanosatellite at 10 to \SI{100}{m} with sub-millimeter accuracy. Advanced missions such as DECIGO and LIFE will also require well-developed technologies for formation flying with three or more spacecraft, along with reliable autonomous control.

To bridge the technology gap towards the future space-based interferometer missions such as DECIGO and LIFE, we propose SILVIA (Space Interferometer Laboratory Voyaging towards Innovative Applications), a small-class mission to demonstrate ultra-precision formation flying. SILVIA will demonstrate sub-micrometer-level precision in three-spacecraft formation flying in low Earth orbit, by orchestrating spacecraft sensors and actuators with laser interferometry and a micro-propulsion system. The low Earth orbit has been known to have economical access to orbit but is dominated by various perturbations compared to those in beyond-Earth orbits. Nevertheless, a recent study \citep{Ito2024} analyzed the near-circular low Earth orbit (LEO) and identified a small relative perturbation region at altitudes above \SI{500}{km}, provided that the spacecraft separation are kept around \SI{100}{m} or less. This region presents an attractive, cost-effective alternative for SILVIA, providing a valuable testing opportunity for building confidence toward future larger-scale missions. The SILVIA mission was first proposed in February 2020 to the call for a mission concept of the competitive middle-class space science mission by the Institute of Space and Astronautical Science of the Japan Aerospace Exploration Agency; then, only the SILVIA mission was promoted to the next development phase among the total seven proposals. SILVIA is currently in the Pre-Phase A2 (mission definition phase), and the next milestone will be the (competitive) final selection process for the middle-class mission.

In this paper, we present the SILVIA mission concept in Section~\ref{sec:MissionConcept}. Section~\ref{sec:Subsystems} reviews key technologies and demonstrates how sub-micrometer precision in relative distance control can be achieved. Section~\ref{sec:Science} explores the scientific potential of 100-meter-class ultra-precision formation flying, particularly in the fields of gravitational wave astronomy and infrared astronomy. Our conclusions and outlook are summarized in Section~\ref{sec:Conclusions}.


\section{SILVIA mission concept} \label{sec:MissionConcept}

The SILVIA mission aims to pave the way for high-precision astronomical observations by formation flying, such as gravitational wave telescopes and optical/infrared interferometers. Its main objective is to demonstrate and mature ultra-precision formation flying technologies in LEO, where economical access to orbit is possible. The mission orbit has been selected as a circular orbit whose altitude is higher than \SI{500}{km} with spacecraft separation of \SI{100}{m}. One reason for this selection is that the separation of \SI{100}{m} or less provides a relatively small perturbation environment in the order of $10^{-7}$ \si{m/s^2} or less, which would be still one order as high as that in beyond-Earth orbits \citep{Ito2024}. Another reason is that proximity formation flying within \SI{100}{m} has been regarded as risky and tended to be avoided \citep{Monnier2019}. Therefore, the selected orbit is still challenging but maintains a good balance between feasibility and risk. In addition, it has the potential to demonstrate the unprecedented precision formation flying within a small-class and experimental program, as pointed out in \citep{Ito2024}. 

Autonomous formation flying by more than two spacecraft has been identified as one of the key technologies for DECIGO- and LIFE-like missions, as they will require managing at least two interferometer arms. Eventually, formation flying with three spacecraft has been chosen as the mission baseline. In addition, a triangular shape has been selected as a primary configuration to demonstrate ultra-precision formation flying, which is exactly the same as that of DECIGO's basic formation and similar in part to that of LIFE's candidate formations (e.g., Emma-X array \citep{Quanz2022}). 

\begin{figure}[t]
 \begin{center}
  \includegraphics[width=0.9\linewidth]{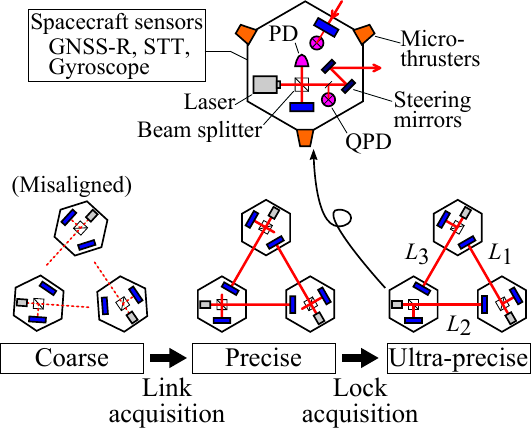} 
 \end{center}
\caption{Schematic of the autonomous procedure to achieve ultra-precise formation flying. Spacecraft and interferometer sensors and actuators we consider in this paper are also shown. GNSS-R: GNSS receiver, STT: star tracker, PD: photodiode, QPD: quadrant PD. }\label{fig:precision}
\end{figure}

The top-level requirement of SILVIA is to suppress the fluctuations of optical paths ($L_1$, $L_2$, $L_3$; see Fig.~\ref{fig:precision}) among each spacecraft within micro- to sub-micrometer order for a specified period, such that
\begin{equation}
    \left |\int^{t_1}_{t_0} \dot{L}_i(\tau) d\tau \right| < \epsilon
    \label{eq:fluctuation}
\end{equation}
where $\epsilon$ $(>0)$ is a very-small fluctuation (e.g., $\epsilon=\SI{1}{\micro.m}$), $\dot{L}_i$ is the time derivative of $L_i$, and $i=1,2,3$. Note that the optical paths are defined based on the optical components (e.g., mirrors), not spacecraft structures. The assumed duration requirement $(t_1-t_0)$ in SILVIA is minimum and relatively-short (e.g., \SI{10}{s}) to provide a highly-stable formation-flying platform for DECIGO-like and LIFE-like missions. A laser interferometer is employed to achieve this unprecedented accuracy in measuring distance fluctuations. In addition, the micro-propulsion system is employed to provide the highly-stable formation-flying platform. 

Figure \ref{fig:precision} shows the autonomous procedure to attain the ultra-precise formation flying. The initial stage of the sequence assumes that the laser interferometers of each spacecraft are misaligned in orbit. In this stage, coarse formation is maintained based on the Global Navigation Satellite System (GNSS)-based relative navigation, which will provide sub-meter to centimeter navigation accuracy. In addition, the spacecraft will use the conventional attitude determination by star-tracker (STT) and gyroscope. In the precise formation sequence, round-trip laser links are established between each spacecraft, allowing continuous monitoring of laser interference signals.
This is achieved by monitoring the beam spot positions on each spacecraft using two-dimensional sensors such as quadrant photodiodes (QPDs) and sharing these measurements via inter-satellite communication. By the end of this sequence, optical path fluctuations are stabilized to the millimeter to sub-millimeter level, limited by the precision of the beam-spot position sensors, using the micro-propulsion system to ensure readiness for transition to the final sequence. The out-of-plane motion is also controlled at the millimeter to sub-millimeter level by using the beam spot sensors. The final sequence is lock acquisition and maintenance, where the laser interferometer fringes are continuously and stably measured and controlled well below one wavelength of the laser. The interference fringe is maintained for a longer duration (e.g., one orbital revolution) than the fluctuation requirement in Eq. (\ref{eq:fluctuation}) to demonstrate long-term stability and reliability.

The autonomous sequence from coarse to ultra-precise formation flying with SILVIA can demonstrate the initial laser link acquisition process for DECIGO and its pre-stabilization, which is necessary for extracting more precise control signals from inter-satellite Fabry-P{\'e}rot cavities. Furthermore, the laser interferometers can measure, test, and verify the in-flight stability of ultra-precise formation flying so that they will be able to provide evidence and confidence with DECIGO- and LIFE-like missions as a flight-proven technology. 

To achieve these goals, it is essential for each spacecraft and their subsystems to work cooperatively. The following section provides an overview of key enabling technologies: laser interferometer, micro thruster, and formation control.

\section{Key technologies} \label{sec:Subsystems}
Achieving the SILVIA mission's goals requires inter-satellite laser interferometers for precise distance and alignment measurements, along with six-degree-of-freedom (6DoF) spacecraft control by micro-thrusters. Maintaining ultra-precise formation of three spacecraft in Earth orbit demands advanced formation flying control, integrating multiple sensors and actuators. In this section, we review each of the key enabling technologies for SILVIA's formation flying. Hereafter, we assume the following configurations and parameters: 
\begin{itemize}
    \item Spacecraft mass is in the order of 100 kg (e.g., 100--\SI{200}{kg}) as a small-class space mission. 
    \item The mission orbit is near-circular in LEO (altitude > \SI{500}{km}). 
    \item The nominal separation ($L_1$, $L_2$, $L_3$ in Fig. \ref{fig:precision}) is \SI{100}{m}. 
    \item The minimum requirement of the mission lifetime is relatively short (e.g., half a year). 
\end{itemize}

\subsection{Laser interferometer}\label{sec:li}
Laser interferometers can achieve precision measurements at length scales much smaller than the wavelength of the laser light, which is on the order of micrometers ~\citep{GRACE-FO}. In the SILVIA mission, laser interferometers are utilized as sensors to precisely control the relative displacement among the spacecraft. As an example, we consider configuring a Michelson interferometer between each pair of the three spacecraft to measure their longitudinal and translational displacements. We will discuss how such high precision can be achieved and the requirements for the laser light source.

As shown in Fig.~\ref{fig:precision}, each Michelson interferometer is asymmetric, with the beam splitter and one end mirror fixed on one spacecraft, and the other end mirror fixed on the other spacecraft. When the length difference between the two arms changes, the phase difference of the reflected light changes, which can be detected as a shift in the interference fringe. In this case, the shorter arm, fixed on one spacecraft, functions as a reference, while changes in the longer arm's length are detected as variations in the distance between spacecraft. When the distance between the spacecraft changes by half a wavelength of the laser, the interference fringe shifts from dark to bright, enabling the measurement of distance fluctuations much smaller than the laser wavelength. The fundamental precision in detecting changes in the interference fringe is limited by quantum fluctuations of photons. The displacement sensitivity limited by this shot noise is given by

\begin{align}
\delta x_\mathrm{s} &= \frac{1}{2\pi}\sqrt{\frac{2hc\lambda_0}{P_0}} \nonumber\\
&\sim 1\times 10^{-15} \left(\frac{\lambda_0}{1.55\,\mathrm{\mu m}}\right)^{1/2}\left(\frac{10\,\mathrm{mW}}{P_0}\right)^{1/2}\,\mathrm{m/\sqrt{Hz}},
\end{align}
where $h$ is the Planck constant, $c$ is the speed of light, $P_0$ is the input laser power, and $\lambda_0$ is the laser wavelength.

To achieve such high displacement sensitivity and linearity in displacement measurements, feedback control is required to stabilize the interference fringe. The interferometer signal is fed back to the thrusters of the spacecraft to minimize the arm length fluctuations. To lock the interferometer using feedback control, the time scale of the control loop must be shorter than that of the error signal. The linear range of the Michelson interferometer error signal is on the order of $\lambda_0$, and it fluctuates at a time scale on the order of $\lambda_0 / v$, where $v$ is the relative speed between spacecraft. For example, this time scale would become around 100 \si{\micro s} or longer if the relative velocity between spacecraft remains up to $v=1$ \si{cm/s} in the coarse formation stage in Fig.~\ref{fig:precision}. On the other hand, the time scale of spacecraft translational control might be in the order of \SI{10}{s} or longer when considering the time response of the thruster and frequency of inter-satellite communication.

To bridge the gap between these time scales, feedback control using laser frequency can be employed, which has been a common technique for terrestrial gravitational-wave detectors. Modulating the laser frequency by $\delta \nu_0$ is equivalent to displacing the end mirror by $\delta L$, with the relationship $\delta L / L = \delta \nu_0 / \nu_0$, where $\nu_0 = c / \lambda_0$ is the laser frequency. The frequency actuator in the laser system has a significantly faster response time (typically below \SI{1}{ms}) than those of the satellite thrusters. Therefore, even if the Michelson interferometer cannot be locked solely through thruster control due to high residual relative velocity between the spacecraft, frequency control can stabilize the interference fringe. By employing hierarchical control, with low frequencies controlled by the thrusters and high frequencies controlled by the laser frequency, long-term lock maintenance can be achieved. Finally, a piezo-actuated mirror provides the fine actuation for the optical path length and further stabilizes the fluctuations down to the requirement in Eq. (\ref{eq:fluctuation}).

For lock acquisition using the laser frequency actuator, the relative velocity between the spacecraft must be reduced to approximately \SI{0.1}{mm/s}. This requirement is determined by the actuator range, which is on the order of \si{GHz}, and the response time scale of spacecraft position control, which is on the order of \SI{10}{s}. Velocity measurement to achieve this deceleration can be performed using several methods. For example, one approach involves counting the number of changes in the interference fringes per unit time, which can then be used to estimate the velocity. Another method is the time-of-flight technique, where the laser intensity is modulated at several tens of megahertz, and the velocity can be estimated from the demodulation phase of the reflected light from the other satellites. Both methods are expected to provide a wide measurement range up to 1 \si{cm/s}, but a detailed feasibility assessment will be part of future work.

Maintaining laser interferometer fringes also requires precise beam pointing control. This can be achieved by using signals from beam spot position sensors, such as QPDs, to provide feedback to the spacecraft thrusters or actuated mirrors for input beam steering. Similar to longitudinal control, a hierarchical control scheme can be employed, where low-frequency components are corrected using thrusters, while high-frequency components are compensated with fast-response steering mirrors. For instance, if beam spot motion on the QPD can be stabilized within 0.1 \si{mm}, the corresponding beam pointing fluctuation remains below 1 \si{\micro rad}, which is sufficient to maintain stable interference fringes in the laser interferometer.

For precise measurements of the distance between spacecraft using laser interferometry, the laser source must be a continuous-wave, linearly polarized laser operating in a single longitudinal and transverse mode. Coherent light sources have been widely used in space for inter-satellite communication and remote sensing lidar. Among available options, the 1.55-\si{\micro m} wavelength range is particularly advantageous due to the extensive development of L-band optical components and the availability of compact, narrow-linewidth lasers.

Three candidates meet the requirements for stability and long coherence length: planar lightwave circuit external cavity diode lasers (PCL-ECL), whispering-gallery-mode (WGM) lasers, and Er-doped fiber distributed-feedback (DFB) lasers. For SILVIA, we adopt the 1.55-\si{\micro m} PCL-ECL (RIO PLANEX) as the baseline laser due to its compactness, narrow linewidth, and high technology readiness level (TRL5). This laser satisfies the fiducial requirements for Michelson interferometer lock acquisition and stability, with an output power exceeding \SI{10}{mW}, a long-term frequency drift below \SI{2}{MHz} over \SI{10}{s}, and a short-term linewidth below \SI{1}{MHz}.

The frequency stability of these lasers is susceptible to external perturbations. The short-term stability is primarily affected by electrical noise, while the long-term stability is dominated by temperature fluctuations of the oscillator. If further stabilization is required, techniques such as frequency locking to an atomic or molecular reference (e.g., Rb two-photon absorption) and active temperature control of the laser system can be implemented to mitigate these effects.

\subsection{Micro thruster}\label{sec:thr}
A low-thrust precise propulsion system is essential to achieve 6DoF precise formation flying. For SILVIA's ultra-precision formation flying, a control resolution of approximately \SI{0.1}{\micro m/s} or lower will be required. On the other hand, a maximum $\Delta V$ (time integral of thrust acceleration) of \SI{10}{cm/s} might be necessary for collision avoidance maneuver of space debris and formation reconfiguration within a reasonable operational duration (e.g., a few hours to one day). To achieve these orbital maneuvers, a maximum thrust in the order of 0.1--\SI{1}{mN} ($10^{-6}$--\SI{e-5}{m/s^2}) will be necessary. In addition, for high-precision astronomical observations, the propulsion system should exhibit low noise, for example, on the order of \SI{1}{\micro N/\sqrt{Hz}}. Developing such a low-thrust, low-noise propulsion system with a wide control range is more demanding than conventional propulsion systems.

There are some possible propulsion methods to satisfy these requirements: Cold-Gas Micro Propulsion (CGMP) system and electric propulsion system such as Field-emission electric propulsion and Electrospray propulsion. Among them, we are currently considering a CGMP system as a primary option for the SILVIA propulsion system. The reasons for this adoption are: (1) the CGMP has flight heritage in overseas missions with a similar requirement to ours \citep[e.g.,][]{noci2009cold}; (2) the CGMP system tends to have smaller power consumption than that of the electric propulsion system. 
In particular, this study assumes a 100-kg-class spacecraft whose power supply capability would be limited, and eight to twelve thrusters will be employed for the 6DoF control of the spacecraft. Therefore, saving the power consumption is one of the key factors. One negative aspect of using CGMP is that its specific impulse is significantly smaller than that of electric propulsion, leading to much faster depletion of propellant. Nevertheless, the CGMP system is still compelling provided that the relatively-short lifetime is acceptable based on the mission-level requirement.

The current baseline of SILVIA's CGMP system mainly consists of a tank, pressure regulating valve, piping, filter, and micro thrusters. Nitrogen will be used as propellant. High-pressure nitrogen gas in the tank will be depressurized by the regulating valve, supplied to the micro thruster, and injected through the thruster nozzle to generate micro thrust force. The main technical challenge lies in the micro thruster. Both throttleable thrusters and pulse-modulated thrusters have been considered a feasible solution, and a trade-off study will be made further. Our preliminary study revealed that the nozzle of the thruster should be fabricated in the order of \si{\micro m} to generate micro-Newton-level thrust, which would be a technical challenge. In addition, micro thrust measurement in a vacuum environment would be another challenge to verify the manufactured micro thrusters. We plan to verify micro-Newton-level thrust control and thrust stability in a ground environment in the breadboard model.

The total amount of propellant will limit the mission lifetime. Our current estimation shows that the total $\Delta V$ of 10 m/s  for each spacecraft will be necessary during a half to one year operation, allocated to coarse/precise formation keeping, reconfiguration, collision avoidance maneuvers, and margins. The micro-propulsion expertise gained from the SILVIA mission will benefit subsequent missions such as DECIGO. Furthermore, it has the potential to contribute to other missions requiring ultra-precision pointing control for astronomical observations.

\subsection{Formation control}
The SILVIA spacecraft formation needs to be controlled under orbital dynamics in LEO. We first review the fundamentals of relative dynamics in LEO, then explain how specific formations are possible under such dynamics. Next, we describe candidate formations during the SILVIA operation. Lastly, we discuss how the coarse to ultra-precise formation control will be attained and tested, in addition to comparing the expected control performance of SILVIA with those of the state-of-the-art missions.  

\begin{figure}
\begin{center}
\includegraphics[width=0.9\linewidth]{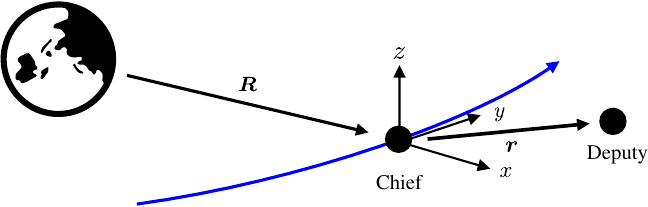}
\end{center}
\caption{Coordinate definition for the relative motion. }
\label{fig:coord}
\end{figure}

First, let us consider two-body dynamics in an unperturbed circular orbit, where two spacecraft (one is called as ``chief'' and another one as ``deputy'') are flying closely with respect to the distance of spacecraft from the center of the main body. Note that the chief can be virtual spacecraft, whereas the deputy denotes real spacecraft. Figure \ref{fig:coord} shows the coordinate definition for the relative motion. In Fig. \ref{fig:coord}, $\boldsymbol{R}$ is the position vector of the chief in the Earth-centered inertial frame and $\boldsymbol{r}=[x,y,z]^T$ is the relative position vector of the deputy with respect to the chief in the local-horizontal local-vertical (LVLH) frame. The LVLH frame is defined as the rotational frame, where the $x$-axis is directed to the radial direction of the chief, $z$-axis is directed to the orbital angular momentum of the chief, and the right-handed frame completes the setup. 

From the linearized dynamics of the deputy with respect to the chief, we obtain a periodic natural solution without thrust and disturbance \citep{Clohessy1960terminal}:
\begin{equation}
\boldsymbol{r}=
\begin{bmatrix}
\rho_x \sin{(\theta+\alpha_x)}\\
\rho_y + 2\rho_x \cos{(\theta+\alpha_x)}\\
\rho_z \sin{(\theta+\alpha_z)}\\
\end{bmatrix}\label{HCW}
\end{equation}
where $\theta$ is the argument of latitude of the chief and $\rho_x$, $\rho_y$, $\rho_z$, $\alpha_x$, and $\alpha_z$ are the (constant) formation parameters. By selecting the appropriate parameters, a natural formation can be designed.

\begin{figure}[t]
\begin{center}
\subfigure[Triangular formation.]{\includegraphics[width=0.9\linewidth]{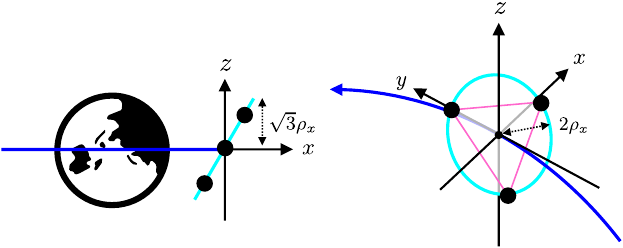}} 
\subfigure[Passively-safe formation.]{\includegraphics[width=0.9\linewidth]{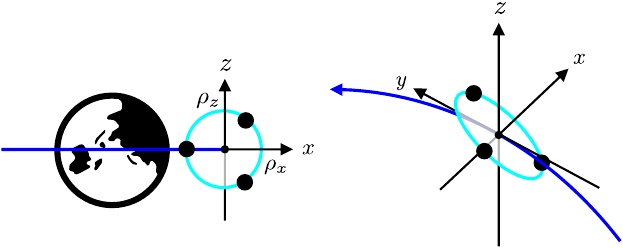}}
\subfigure[Linear formation (optional).]{\includegraphics[width=0.9\linewidth]{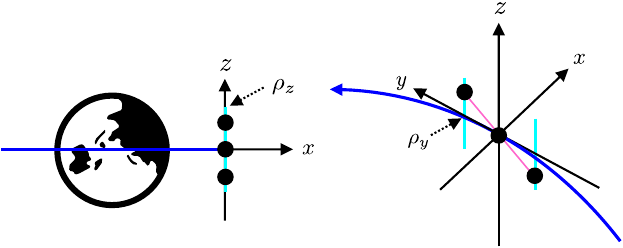}}
\end{center}
\caption{Formation examples for (a) triangular, (b) passively-safe, and (c) linear formations. The blue, light-blue, and magenta lines show the reference orbits of the chief, periodic relative orbits of the deputies, and optical paths, respectively. }
\label{fig:ff}
\end{figure}

\begin{table*}
    \centering
    \caption{Main usage and description for the three natural formations.}
    \label{tab:formation}
    \begin{tabular}{lll}
         \toprule
         Formation & Main usage & Description on natural formation along Eq. (\ref{HCW}) \\  \midrule
         (a) Triangular & Maintenance of laser interference fringe &  Distance: constant ($L_i=2\sqrt{3}\rho_x$, $i=1,2,3$). \\
         (b) Passively-safe & Initial commissioning \& abort & Secured minimum separation (collision-free). \\ 
         (c) Linear (optional) & Maintenance of stellar interference fringe & Alignment: orthogonal to a target direction, OPD: 0.  \\ \bottomrule
    \end{tabular}
\end{table*}

Figure~\ref{fig:ff} shows the formation examples in the SILVIA operation: (a) triangular, (b) passively-safe, and (c) linear formations and their main usage and description on natural formation along Eq. (\ref{HCW}) are summarized in Table~\ref{tab:formation}. The triangular formation is the primary formation to demonstrate the coarse to ultra-precise formation flying as shown in Section \ref{sec:MissionConcept}. It is realized on a general circular orbit (GCO) \citep{Alfriend2010spacecraft} by selecting $\rho_y=0$, $\rho_z=\sqrt{3}\rho_x$, $\alpha_z=\alpha_x$, and shifting $\alpha_x$ by $(2\pi)/3$ among three spacecraft (e.g., $\alpha_x=0, (2\pi)/3, (4\pi)/3$). As a result, the relative distances $L_i$ $(i=1,2,3)$ in Fig. \ref{fig:precision} are naturally maintained at constant ($L_i=2\sqrt{3}\rho_x$).
A passively-safe formation is realized by selecting e.g., $\rho_y=0$, $\rho_z=\rho_x$, $\alpha_z=\alpha_x\pm\pi/2$, and shifting $\alpha_x$ by $(2\pi)/3$ among three spacecraft. It is well known that the along-track ($y$) axis has large uncertainty against disturbance compared to the other axes \citep{d2010autonomous}. The passively-safe formation can avoid encountering the close relative positions of spacecraft in the $(x-z)$ plane, and this is achieved by separating $\alpha_x$ and $\alpha_z$ by (close to) $\pi/2$. It will be used for the initial commissioning phase before the autonomous formation-flying functions are verified in orbit. Another example for use will be an abort situation when each spacecraft is at a high risk for collision, and need to transfer to a safer orbit. 
The last formation is the linear one, where one spacecraft is placed at the origin and the other two spacecraft are placed symmetrically at $\rho_x=0$, $\rho_y=\pm L$, $\rho_z=\rho_y\tan{p}$, and $\alpha_z=q$. The parameters $p$ and $q$ are relevant to the observation direction and chief's orbital element \citep{hansen_ireland_2020}. This natural formation can align orthogonally to the target direction in the inertial frame. Furthermore, it can maintain the zero OPD while each optical path length varies between $L$ and $L\sqrt{1+\tan^2{p}}$. Therefore, it is suitable for observing a target star as a formation-flying astronomical interferometer. This formation might be used if the SILVIA spacecraft can additionally employ an astronomical interferometer, but we need a further study on its feasibility.

These formations derived from Eq. (\ref{HCW}) are theoretically maintained without control, under the linearized dynamics in an unperturbed circular orbit. However, the relative motions can be perturbed by various disturbances in LEO such as the Earth zonal $J_2$ gravitational potential, non-zero mean eccentricity, nonlinear terms that are neglected in the linearization process to obtain Eq. (\ref{HCW}), atmospheric drag, and solar radiation; these relative perturbations must be compensated by thrusters. {During the ultra-precision formation flying, the spacecraft have to compensate for all of the perturbing forces comprising short-periodic, long-periodic, and secular ones,} and its total estimated magnitude is in the order of sub-\si{\micro m/s^2} with a separation of 100 m at an altitude of 500--600 km \citep{Ito2024}. This compensation would be feasible with the micro-propulsion system in Section~\ref{sec:thr}. The coarse formation flying maintains its mean relative orbit by compensating for secular perturbations  only, thereby the propellant consumption rate will be lower compared to the precise formation modes. 

The SILVIA mission achieves ultra-precision formation flying through cooperative control of spacecraft via inter-satellite communication. The transition from coarse to ultra-precise formation is critical in the SILVIA mission. In the coarse formation phase, the formation accuracy will be limited by GNSS-based navigation. The PRISMA mission achieved GNSS-based onboard relative navigation accuracy of below 10 cm and 0.5 mm/s (three-dimensional, root-mean-square) for relative position and velocity in most of the operational scenarios \citep{d2012spaceborne}. The SILVIA mission aims to extend it to three spacecraft navigation while maintaining the same level of accuracy of PRISMA. The precise and ultra-precise formation stages in Fig. \ref{fig:precision} are more unique to the SILVIA mission, finally stabilizing the optical paths within micro-to sub-micrometer order for a specified time period. This procedure involves cooperative control using spacecraft thrusters and laser interferometer actuators. Additionally, onboard navigation must achieve the highest possible precision by integrating GNSS-based measurements, beam position sensors (QPDs), relative velocity and displacement measurements along the laser beam axis, while also incorporating real-time interferometer feedback signals to the laser frequency and steering mirrors. To refine this integrated navigation and control, we are now developing a unique hardware-in-the-loop testbed for ultra-precision formation flying \citep{iwaki2}. We plan to test a prototype algorithm from coarse to ultra-precise formation control on this testbed with a breadboard model of a laser interferometer. 

Compared to the recent ongoing missions such as PROBA-3 and VISORS, the unique control requirement of the SILVIA mission is to maintain the optical paths between spacecraft with the ultra-high precision (e.g., $\epsilon=\SI{1}{\micro.m}$ in Eq. (\ref{eq:fluctuation}) for 10 seconds and the maintenance of the laser interference fringe for one orbital revolution). To do this, each spacecraft of SILVIA will employ the sub-milli-Newton-class micro thrusters in Section \ref{sec:thr}, which are one-to two orders of magnitude as small as those in the occulter spacecraft of PROBA-3 (10 mN for each thruster \citep{Penin2020proba3}) and chief/deputy spacecraft of VISORS (20 mN for each \citep{Lightsey2018cgj,Koenig2023visors}). In addition, the SILVIA spacecraft will utilize the ultra-precise measurements of the laser interferometer (science instrument) for onboard navigation, which also contrasts with the performance of the relevant missions. Reliable and autonomous navigation and control with the unprecedentedly precise actuators/sensors are one of the major technical challenges in SILVIA's formation flying demonstration.  

\section{Science with 100-meter-class ultra-precision formation flying} \label{sec:Science}
While full-scale missions like DECIGO and LIFE are essential for astronomical observations, equipping the 100-meter-class SILVIA platform with scientific instruments could enable important in-orbit demonstrations for gravitational wave astronomy and infrared astronomy. Here, we explore the scientific opportunities achievable even with 100-meter-class ultra-precision formation flying.

\subsection{Demonstration of gravitational wave searches in orbit}
\begin{figure}[t]
 \begin{center}
  \includegraphics[width=0.9\linewidth]{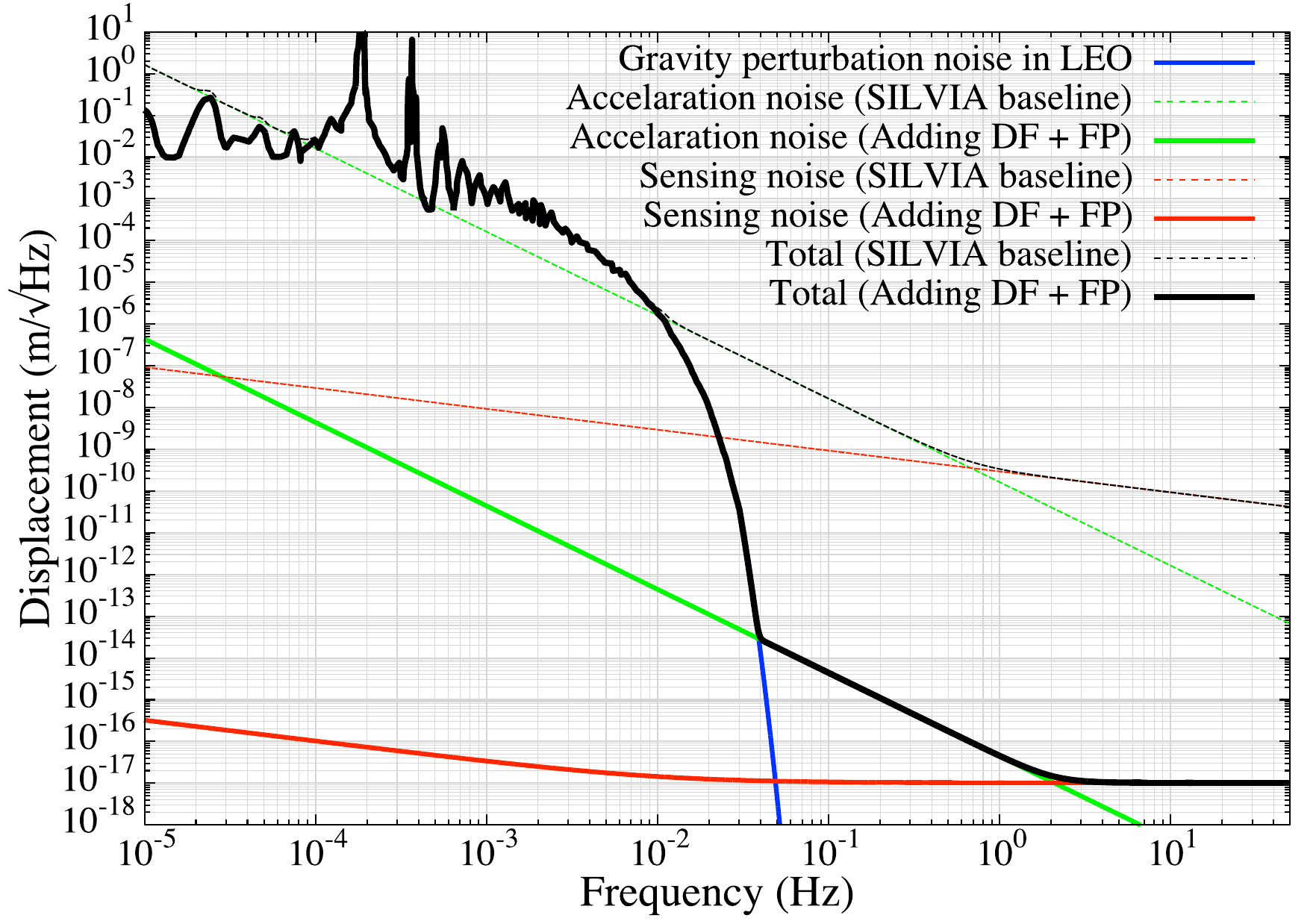} 
 \end{center}
\caption{Estimated displacement sensitivity of SILVIA with potential noise sources. The plotted noise contributions include gravitational fluctuations (blue), acceleration noise (green), and sensing noise (red). Dotted lines represent the SILVIA baseline, while solid lines correspond to SILVIA with drag-free (DF) control and Fabry-P{\'e}rot (FP) cavities between spacecraft, as described in the main text. The gravitational fluctuation noise spectrum shown here is derived from GRACE-FO measurements~\citep{GRACE-FO}, scaled to match the 100-meter spacecraft separation of SILVIA. }
\label{fig:SILVIAnoise}
\end{figure}

\begin{figure}[t]
 \begin{center}
  \includegraphics[width=0.9\linewidth]{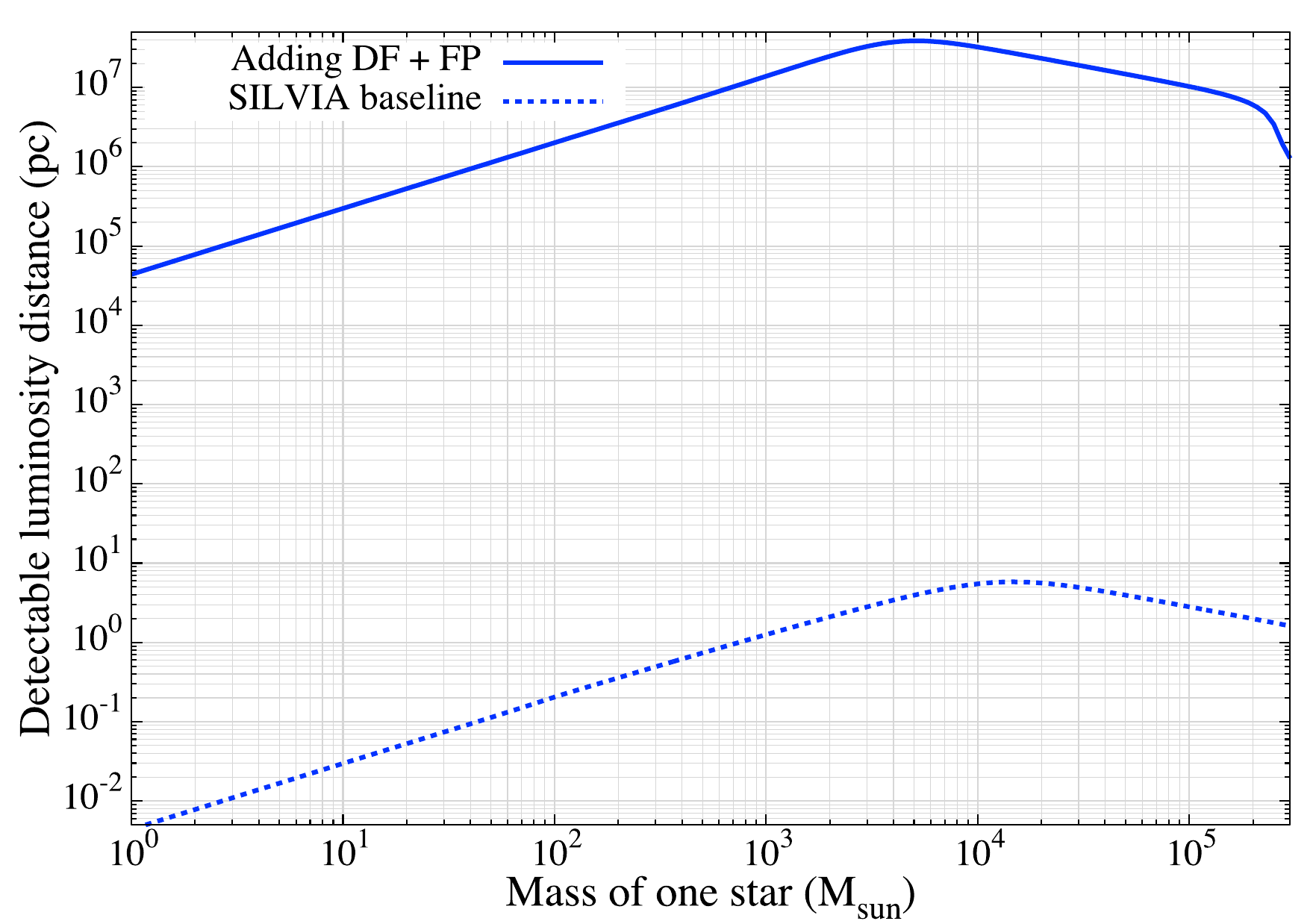} 
 \end{center}
\caption{Detection ranges of SILVIA as a gravitational wave detector. The sky- and polarization-averaged luminosity distance at which gravitational waves from an equal-mass, non-spinning compact binary coalescence can be detected with a signal-to-noise ratio of 8 is shown. Dotted lines represent the SILVIA baseline, while solid lines correspond to SILVIA with drag-free control and Fabry-P{\'e}rot cavities between spacecraft, as described in the main text. }
\label{fig:GWrange}
\end{figure}

In its baseline configuration, SILVIA does not include floating test masses. However, by treating the spacecraft themselves as test masses, the fundamental principles of gravitational wave detection can be demonstrated without additional instruments. In this scenario, thruster noise and solar radiation pressure fluctuations introduce displacement noises at low frequencies. Additionally, with an asymmetric Michelson interferometer, frequency noise contributes to sensing noise at higher frequencies. For instance, assuming a frequency noise of $5 \times 10^2$~\si{Hz/\sqrt{Hz}}, a thruster noise level of 1~\si{\micro N/\sqrt{Hz}}, and spacecraft mass of 200~\si{kg}, the displacement sensitivity at 1~\si{Hz} could reach $4\times 10^{-10}$~\si{m/\sqrt{Hz}} (see Fig.~\ref{fig:SILVIAnoise}). This sensitivity would enable searches for $10^{4} M_{\rm \odot}$--$10^{4} M_{\rm \odot}$ binary black hole mergers for up to 5\,\si{pc} (see Fig.~\ref{fig:GWrange}).

If SILVIA were equipped with floating test masses and employed drag-free control~\citep{LISAPathfinder, GRATTIS}, external disturbance noise could be further suppressed. However, in LEO, variations in Earth's gravity field would still limit the low-frequency sensitivity. Additionally, if Fabry-P{\'e}rot cavities were implemented between spacecraft, as in DECIGO, various sensing noises could be reduced, ultimately achieving a sensitivity limited by shot noise. Assuming an acceleration noise level to that of LISA Pathfinder, a cavity finesse of 100, and an input power of 10~\si{mW}, the displacement sensitivity at 1~\si{Hz} could reach the $10^{-17}$~\si{m/\sqrt{Hz}} level. This would enable searching for $10^{4} M_{\rm \odot}$--$10^{4} M_{\rm \odot}$ binary black hole mergers for up to 30~\si{Mpc}. Even under this optimistic scenario, gravitational wave detection remains challenging, due to limited arm length. However, a platform provided by SILVIA would provide unprecedented sensitivity for intermediate-mass black hole binaries, making it a critical technology demonstration for future missions.

\subsection{Demonstration of nulling interferometry in orbit}
In this section, we discuss the possibility of testing nulling interferometry in orbit using a highly stable formation-flying platform such as SILVIA, potentially paving the way for future LIFE-like missions. 
An important and relevant mission to SILVIA is SEIRIOS \citep{IkariEtc21, Matsuo2022}, which was mentioned in Section \ref{sec:Intro}. In comparison to SEIRIOS, SILVIA aims to demonstrate a high stability of micron-to-sub-micron order between the laser-interferometric optical paths of spacecraft for a specified period, as described in Section \ref{sec:MissionConcept}. This stable platform might be suitable for on-orbit testing of more advanced interferometric technologies than those planned for SEIRIOS. 

From a programmatic standpoint, the stability of SILVIA might be more compatible with the LIFE-like missions; it has the potential of testing high-precision OPD compensation in orbit for nulling interferometry. With this background, we consider nulling interferometry as an optional scientific demonstration. Two collector spacecraft will be equipped with a flat mirror to collect light from celestial objects and direct it to the beam combiner spacecraft, where the two incoming beams are combined. The beam combiner will include the following subsystems: 1) a compressor to reduce the beam diameter, 2) a tip-tilt system to measure and correct for the relative tilt between the two beams sent from the collector spacecraft, 3) a delay line system to compensate for the OPD, 4) a fringe-tracking system to measure the OPD, and 5) a beam combiner and science camera for observing the resulting interference fringes.

In the SILVIA demonstration, the initial accuracy of the relative position between spacecraft (i.e., coarse phase in Fig. \ref{fig:precision} left) will be on the order of sub-meter to centimeter by GNSS navigation. Thus, the OPD measurement and compensation must be refined incrementally using the fringe-tracking system and science camera. The coarse OPD measurement could be achieved by utilizing state-of-the-art instrumentation techniques such as fringe measurement based on densified pupil technique \citep{Matsuo+2016, Matsuo2022} , which can extend the coherence length to around 1 mm. A more precise OPD measurement could be obtained from the constructed and nulled fringes from the science camera. Once the OPD is within one-tenth of the observing wavelength, the null depth will become highly sensitive to OPD variations. Under these conditions, real-time and precise OPD measurement and control will be achieved using a science camera and a delay line system.   

\begin{figure*}
 \begin{center}
  \includegraphics[width=13cm]{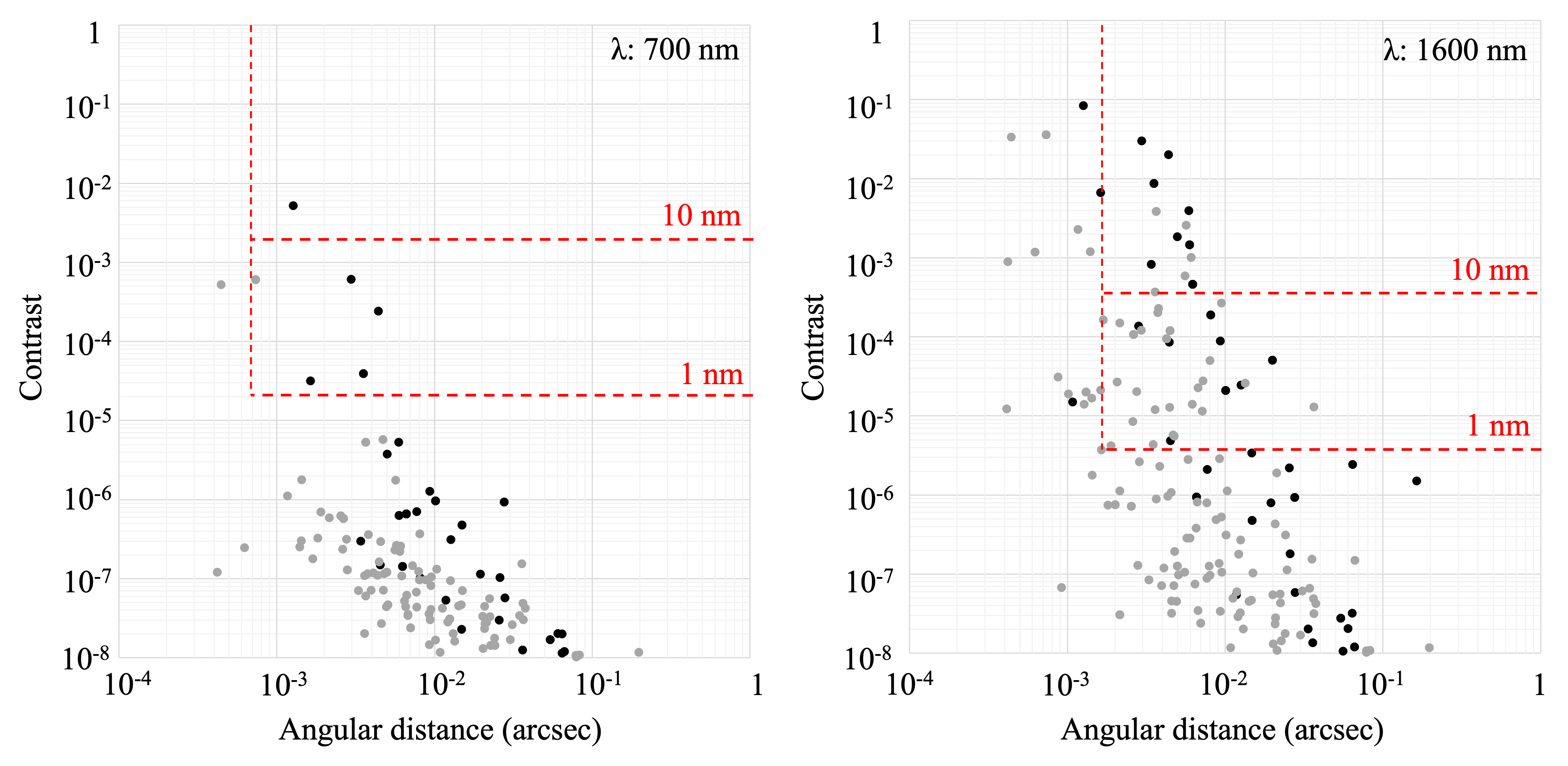} 
 \end{center}
\caption{Angular distance versus contrast at 700 nm (left) and 1600 nm (right) for exoplanets within 20 pc that have measured planetary radii and semi-major axes. All samples were taken from the list in Exoplanet Encyclopedia\protect\footnotemark[1]. Black dots represent exoplanets whose angular separations are listed in the list, while gray dots indicate those whose angular separations are estimated from their semi-major axes and the distances to the host star. The contrast ratio between the total planetary flux (reflected light + thermal emission) and the host star is given by $p \phi(r_{p}/a)^2 + B_{\lambda}(T_{p})r_{p}^{2}/B_{\lambda}(T_{s})r_s^{2}$, where $p$ is the geometric albedo, $\phi$ is the phase function, $a$ is the semi-major axis, $r$, $T$, and $B_{\lambda}$ represent the radius, temperature, and blackbody radiation at wavelength $\lambda$, respectively; subscripts $p$ and $s$ refer to the planet and its host star, respectively. For this plot, we adopt $p = 0.15$ (the lowest upper limit reported for hot Jupiters from \cite{Blavzek+2022}), and $\phi = 1/\pi$. Planetary radii were estimated based on the mass-radius relation from \citep{Muller+2024}. Planet temperatures were simply calculated assuming thermal equilibrium. The inner working angle of nulling interferometer was set to 0.7 and 1.5 milliarcseconds at 0.7 and 1.5 \SI{}{\micro m}, corresponding to 0.5 $\lambda / B$ for a baseline length of \SI{100}{m}. The contrast limit in visible and near-infrared, determined by the optical path difference (OPD), $h$, is calculated as $(\pi h / \lambda)^2$. The contrast curves are shown for the OPD values of 1nm and 10nm, respectively.  }\label{fig:IRscience}
\end{figure*}

If an additional stellar beam combiner is implemented, it can theoretically achieve a spatial resolution of 0.7 and 1.6 milliarcseconds in visible (0.7 \SI{}{\micro m}) and near-infrared (1.6 \SI{}{\micro m}), respectively. This spatial resolution is derived from the relationship of 0.5 $\lambda/B$, where $B$ is the baseline length of SILVIA. In addition, its highly stable platform has the potential to attain deeper null depths through nulling interferometry. Combining extremely high spatial resolution with moderate contrast might allow the direct detection of faint objects situated very close to their host stars. Figure~\ref{fig:IRscience} shows the angular distance versus contrast for exoplanets within 20 pc that have measured planetary masses and semi-major axes. As shown in Fig.~\ref{fig:IRscience}, the deep null depth, reaching down to $10^{-5}$, may allow for the direct detection of the total planetary flux (reflected light + thermal emission) from hot Jovian planets orbiting very close to their host stars. Although observing at longer wavelengths may introduce additional complexity, such as the need for cooling, it provides two key advantages compared to visible wavelengths: (1) the potential for deeper contrast, and (2) increased thermal emission from exoplanets, resulting in a lower contrast ratio. Therefore, the possibility of testing nulling interferometry in orbit is considered not only in the visible but also in the near-infrared. This capability would be invaluable for future high-contrast imaging missions, such as the Habitable Worlds Observatory (HWO) \citep{Lee2024HWO} and LIFE, particularly from the perspective of target selection. 

\footnotetext[1]{https://www.exoplanet.eu}

\section{Conclusions} \label{sec:Conclusions}
In this paper, we proposed the SILVIA mission concept for demonstrating ultra-precision formation flying between three spacecraft. Through a review of previous studies, we found that while the highest formation flying accuracy demonstrated in orbit so far has reached the centimeter-to sub-millimeter level, future space-based interferometers such as DECIGO and LIFE will require sub-micrometer precision in controlling optical paths between spacecraft, by integrating sensors and actuators for spacecraft and interferometers. To demonstrate this level of precision in a cost-effective manner, we identified a 100-meter-scale mission in a near-circular low Earth orbit as an optimal approach, as this configuration minimizes both acceleration perturbations and the risk of collision.

We showed that such precision could be achieved by employing a control strategy that progressively integrates multiple ranging sensors with different measurement ranges and accuracies, from GNSS to interferometry between spacecraft, while utilizing micro-thrusters and interferometric actuators for precise feedback control. Integrating scientific instruments into the SILVIA platform could facilitate the first in-orbit search for gravitational waves and the demonstration of nulling interferometry, though with limited sensitivity. This mission is expected to pave the way for larger-scale projects such as DECIGO and LIFE while also offering applications in a wide range of fields.

SILVIA is currently in the Pre-Phase A2 (mission definition phase) with the support of the Institute of Space and Astronautical Science, and is being studied for a target launch in the early 2030s.

\begin{ack}

This work was conducted in the formation flight working group (which began in 2006) led by the Japan Aerospace Exploration Agency. This work was supported by the Advisory Committee for Space Engineering and the Institute of Space and Astronautical Science. We thank Naoki Kohara of the National Astronomical Observatory of Japan for giving valuable advice on the conceptual design of laser interferometers. 
\end{ack}

\bibliographystyle{apj}
\bibliography{references.bib}
\end{document}